
\input harvmac.tex
\input tables

\def\aa{{\alpha}}
\def\bb{{\beta}}
\def\cc{{\gamma}}
\def\dd{{\delta}}

\def\va{{\triangle \aa\over \aa}}
\def\vb{{\triangle \bb\over \bb}}
\def\vc{{\triangle \cc\over \cc}}
\def\vd{{\triangle \dd\over \dd}}

\def\IR{\relax{\rm I\kern-.18em R}}
\def\IZ{\relax\ifmmode\mathchoice
{\hbox{\cmss Z\kern-.4em Z}}{\hbox{\cmss Z\kern-.4em Z}}
{\lower.9pt\hbox{\cmsss Z\kern-.4em Z}} {\lower1.2pt\hbox{\cmsss
Z\kern-.4em Z}}\else{\cmss Z\kern-.4em Z}\fi}
\font\cmss=cmss10 \font\cmsss=cmss10 at 7pt

%

%



\lref\vann{P. Van Nieuwenhuizen, {\sl The Complete Mass Spectrum of 
d=11 Supergravity Compactified on $S^4$ and a General Mass Formula 
for Arbitrary Cosets $M_4$ }, Class. Quantum. Grav. {\bf 2} (1985) 1.}

\lref\juan{J. Maldacena, {\sl The Large N Limit of Superconformal Field 
Theories and Supergravity}, {\tt hep-th/9711200}. }

\lref\witten{E. Witten, {\sl Anti de Sitter Space and Holography}, 
{\tt hep-th/9802150}.}

\lref\igoste{S.S. Gubser, I. Klebanov and A. Polyakov, {\sl Gauge 
Theories Correlators from Non-critical String Theory}, {\tt
hep-th/9802190}.  }

\lref\shamiteva{S. Kachru and E. Silverstein, {\sl 4D Conformal Field 
Theories and Strings on Orbifolds}, Phys. Rev. Lett. {\bf 80} (1998) 4855.}

\lref\banks{T. Banks and Z. Zaks, {\sl On the Phase Structure of 
Vectorlike Gauge Theories with Massless Fermions}, Nucl. Phys. 
{\bf B196} (1982) 189. }

\lref\vafaber{ A. Lawrence, N. Nekrasov and C. Vafa, {\sl On 
Conformal Field Theories in Four Dimensions}, {\tt hep-th/9803015};
\quad
M. Bershadsky, Z. Kakushadze and C. Vafa, {\sl String Expansion as
Large N Expansion of Gauge Theories}, Nucl. Phys. {\bf B523} 
(1998) 59.}

\lref\evshamb{S. Kachru, J. Kumar and E. Silverstein, {\sl Vacuum Energy 
Cancellation in a Non-supersymmetric String}, {\tt hep-th/9807076}.}

\lref\freundrubin{
P.G.O Freund and M.A. Rubin, {\sl Dynamics of Dimensional
Reduction}, Phys. Lett. {\bf 97B} (1980) 223. }

\lref\castellaniromanswarner{
L. Castellani, L.J. Romans and N.P. Warner, 
{\sl Classification of Compactifying Solutions of $D=11$ Supergravity}, 
Nucl. Phys. {\bf B241} (1984) 429.}

\lref\other{M.J. Duff and D.J. Toms, in the 2nd Europhysical Study Conf. 
in Unification, (Erice, 1991); \quad
E. Witten, {\sl Search for A Realistic Kaluza-Klein Theory}, 
Nucl. Phys. {\bf B186} (1981) 412}

\lref\casta{L. Castellani, R. D'Auria and P. Fre, {\sl $SU(3) \times 
SU(2) \times U(1)$ from $D=11$ Supergravity}, Nucl. Phys. {\bf B239}
(1984) 610; L. Castellani and L.J. Romans, {\sl N=3 and N=1
Supersymmetry in a New Class of Solutions for d=11 Supergravity},
Nucl. Phys. {\bf B238 } (1984) 683.}

\lref\castb{L. Castellani, R. D'Auria, P. Fre, K. Pilch and 
P. van Nieuwenhuizen, {\sl The Bosonic Mass Formula for Freund-Rubin 
Solutions of $D=11$ Supergravity on General Coset Manifolds}, 
Class. Quant. Grav. {\bf 1} (1984) 339.}

\lref\dineseiberg{M. Dine and N. Seiberg, {\sl Is The Superstring
Weakly Coupled?}, Phys. Lett. {\bf 162B} (1985) 299.}

\lref\ubound{P. Breitenlohner and D.Z. Freedman, Ann. Phys.
{\bf 144} (1982) 249; L. Mezincescu and P.K. Townsend, Ann. Phys. 
{\bf 160}
(1985) 406.}

\lref\evadine{M. Dine and E. Silverstein, {\sl New M-Theory Backgrounds
with Frozen Moduli}, {\tt hep-th/9712166}.}

\lref\nahm{W. Nahm, {\sl Supersymmetries and Their Representations},
Nucl. Phys. {\bf B135} (1978) 149.}

\lref\besse{A.L. Besse, {\sl Einstein Manifolds}, (1980, Springer-Verlag,
New York).}

\lref\egh{T. Eguchi, P.B. Gilkey and A.J. Hanson, Phys. Rep. {\bf 66} 
(1980) 213. }

\lref\adsseven{M. Berkooz, {\sl A Supergravity Dual of a $(1,0)$ 
Field Theory in Six-Dimensions}, {\tt hep-th/9802195}; \quad
O. Aharony, Y. Oz and Z. Yin, {\sl M-Theory in 
$AdS(P)\times S(11-P)$ and Superconformal Field Theories}, 
{\tt hep-th/9803051}; \quad
S. Minwalla, {\sl Particles on AdS(4/7) and Primary Operators on M(2) 
and M(5)-brane World Volume}, {\tt hep-th/9803053}; \quad
R.G. Leigh and M. Rozali, {\sl The Large N limit of the 
(2,0) Superconformal Field Theory}, {\tt hep-th/9803068}; \quad
E. Halyo, {\sl Supergravity on $AdS(4/7) \times S(7/4)$ and M-Branes},
J. High Energy Phys. {\bf 4} (1998) 11; \quad
S. Ferrara, A. Kehagias, H. Partouche and A. Zaffaroni, {\sl Membranes
and Five-Branes with Lower Supersymmetry and Their AdS Supergravity
Duals}, {\tt hep-th/9803109}; \quad
C. Ahn, K. Oh and R. Tatar, {\sl Orbifolds of $AdS_7 \times S_4$
and Six-Dimensional $(0,1)$ SCFT}, {\tt hep-th/9804093}.}

\lref\adsfour{O. Aharony, Y. Oz and Z. Yin, {\sl M-Theory in 
$AdS(P)\times S(11-P)$ and Superconformal Field Theories}, 
{\tt hep-th/9803051}; \quad J. Gomis. {\sl Anti-de-Sitter Geometry and 
Strongly Coupled 
String Theory}, {\tt hep-th/9803119}; \quad
E. Halyo, {\sl Supergravity on $AdS(5/4) \times$ Hope Fibration and
Conformal Field Theories}, {\tt hep-th/9803193}; \quad
R. Entin and J. Gomis, {\sl Spectrum of Chiral Operators in Strongly
Coupled Gauge Theories}, {\tt hep-th/9804060}.}

\lref\duffreview{M.J. Duff, B.E.W. Nilsson and C.N. Pope, 
Phys. Rep. {\bf 130} (1986) 1.}

\lref\dnp{M.J. Duff, B.E.W. Nilsson and C.N. Pope,
{\sl The Criterion for Vacuum Stability in in Kaluza-Klein Supergravity},
Phys. Lett. {\bf 139B} (1984) 154}

\lref\bf{P. Breitenlohner adn D.Z.Freedman, {\sl Stability in Gauged 
Extended Supergravity}, Ann. Phys. 144 (1982) 173.}

\lref\wtntor{E. Witten, {\sl Baryons and Branes in Anto-De-Sitter Space},
J.High Energy Phys. {\bf 98-07} (1998) 6.}

\lref\squash{M.J. Duff, B.E.W. Nilsson and C.N. Pope, {\sl Spontaneous 
supersymmetry breaking by the squashed seven-sphere}, Phys. Rev. Lett. 50
(1983) 2043, Erratum 51 (1983) 846.}

\lref\savetc{S. Sethi, {\sl A Relation Between N=8 Gauge Theories in 
Three Dimensions}, hep-th/9809162; M. Berkooz and A. Kapustin {\sl New 
IR Dualities in Supersymmetric Gauge Theories in Three Dimensions}, 
hep-th/9810257; C. Ahn, H. Kim, B.-H. Lee, H.-S. Yang, {\sl N=8 SCFT 
and M-theory on $AdS(4)\times RP^7$}, hep-th/9811010}


\Title{\vbox{\baselineskip12pt\hbox{\tt hep-th/9807200}
\hbox{IASSNS-HEP-98/68} \hbox{SNUTP 98/088}}}
{\vbox{\centerline{Non-Supersymmetric Stable Vacua of M-Theory} }}

\centerline{Micha Berkooz$^1$ and Soo-Jong Rey$^2$} 
\bigskip
\bigskip

\centerline{\sl $^1$ School of Natural Sciences, Institute for 
                     Advanced Study}
\centerline{\sl Princeton, NJ 08540 USA}
\smallskip
\smallskip
\centerline{\sl $^2$ Physics Department, Seoul National University}
\centerline{\sl Seoul 151-742 KOREA}
\smallskip
\smallskip
\centerline{\tt berkooz@sns.ias.edu, sjrey@gravity.snu.ac.kr}
\bigskip
\bigskip
\bigskip
\noindent
\centerline{\bf abstract}

We discuss the stability of non-supersymmetric compactifications of
M-theory and string theory of the form $AdS\times X$, and their dual
non-supersymmetric interacting conformal field theories. We argue that
some of the difficulties in controlling $1/N$-corrections disappear in
the cases that the large-N dual conformal field theory has no
invariant marginal operators (and in some cases with no exactly
marginal operators only).  We provide several examples of such
compactifications of M-theory down to $AdS_4$.

\Date{July 1998}

\newsec{Introduction}

A powerful tool in the 
study of large N limits of conformal field
theories is their description in terms of certain string
theory or M-theory vacua
\juan. These vacua are of the form $AdS_{p}\times X_{q}$, where $X_q$
is a compact Einstein manifold and $p+q$ adds up to 10 or 11. A
precise prescription of how to derive conformal dimensions and 
correlation functions from the supergravity side was given in 
\witten, \igoste\ and has been extensively developed since.

Of particular interest is the description of vacua that break
supersymmetry completely. The interest is drawn from both field
theoretic and M-theory points of view. From the field theoretic
point of view, one would like to have some control over
non-supersymmetric interacting conformal field theories as a tool for
better understanding non-supersymmetric field theories in general. The
existence of such conformal fixed points is well established \banks\
but only in the perturbative regime. There is no concrete proposal for a
strongly coupled non-supersymmetric conformal field theory in $d > 2$.

The issue is even more interesting from the point of view of 
M-theory. The supergravity background that is dual to the large N 
limit of such conformal field theory will be a vacuum with no 
supersymmetry which is nevertheless stable. 
The understanding of stable non-supersymmetric vacua 
would be an important step forward 
in order for string theories to make contact with reality.

A procedure for obtaining non-supersymmetric supergravity vacua that
may be dual to $d=4$ non-supersymmetric conformal field theories has
been discussed in \shamiteva {} in the context of Type IIB string
theory orbifolds (related issues have been studied subsequently in
\vafaber).  As discussed in that work, there might be a non-vanishing
dilaton tadpole, which would destabilize the vacuum (unless the
dilaton potential remains flat, as was suggested recently \evshamb {}
for some cases).  This implies that the true quantum vacuum may be
quite far away in the space of vacua. One indeed expects that this
would be a generic problem of orbifolds at weak string coupling. The
phenomenon is the well-known Dine-Seiberg problem re-emerging in this
context \dineseiberg.

In this paper, we will discuss a construction in M-theory of vacua
which break supersymmetry completely, but nevertheless overcome this 
specific problem: there is no
candidate for a field which might develop a significant tadpole when
taking into account possible ${1\over N}$ corrections (although we do
not know how to actually compute them in M-theory). The usefulness of
M-theory for freezing moduli was already observed in \evadine. We will
focus primarily on compactification down to $AdS_4$.

In the following section, we will discuss an argument why a certain
class of vacua of M-theory may be close to a stable one in a sense
that will be made precise there.  In section 3 we discuss a concrete
set of non-supersymmetric vacua of M-theory of the form
$AdS_4$ times an Einstein manifold $M_7$ which fall into this class.

 
\newsec{A Conformal Field Theory Argument for Stability}

Suppose we are given an M-theory vacuum which is dual, as
$N\rightarrow\infty$, to a conformal field theory. One can then argue
that if the conformal field theory contains no invariant marginal
operators then the effects of the tadpoles is to modify the true
conformal fixed point as well as the true supergravity vacuum, whether
supersymmetric or not, by effects that are only of order ${1 / N}$.

The main assumption that goes into the argument is what we will call
{\sl calculability}. We will assume that given a supergravity vacuum
at large $N$ limit there exists a well-defined calculational procedure
such that this vacuum corresponds to the classical theory. This is a
very natural assumption from the point of view of supergravity side
but when translated to the conformal field theory side it becomes very
powerful. 

Since we do not know how to quantize such vacua of M-theory or string
theories with Ramond-Ramond flux, we cannot compute the $1\over N$
corrections. Nevertheless, the assumption of calculability can carry
us some steps forward. The implication of this assumption is that
gauge symmetries of the classical supergravity background can be
broken only spontaneously, if at all, by the ${1\over N}$
corrections. In that case, any supergravity field that is charged
under the gauge symmetries would not have a linear tadpole (of the
form ${\cal L}_{\rm SUGRA} = \cdots +
\lambda^* \phi + \lambda \phi^* +  \cdots$) 
since such a tadpole will necessarily break gauge symmetries
explicitly. Since we cannot generate a linear tadpole for the charged
fields, we can at most change their mass term (by $1/N$
corrections). This in turn will change their dimension around the
conformal fixed point but will not destabilize it.  Higher order
corrections to these fields will not even change the dimension. Our
only concern, therefore, are linear tadpoles for fields that are {\sl
neutral} under all the global symmetries of the dual conformal field
theory, (i.e., the associated particles are not charged under the
gauge symmetries of the classical supergravity background). If the
field has non-zero mass, viz., it satisfies the equation of motion
$(\Delta_{AdS}+m^2 ) \phi=0$ with $m\not=0$, then a tadpole of the form
${\cal L}_{\rm SUGRA} = {\cal L}_0 + {C \over N} \phi + \cdots$ in the
Lagrangian will only shift the expectation value of $\phi$ by an
amount proportional to $1 \over N$ but otherwise the solution will
still retain the $AdS$ symmetries. This will not be true for a
massless field where now the equation of motion is $\Delta\phi+{C
\over N} =0$. One concludes that if there are no massless invariant
scalar fields, then any $1 \over N$-corrections to the equations of
motion does not destabilize the solution

A parallel argument can be made in the terms of the field theory. The
large N limit of a given quantum field theory is defined in terms of
rescaled couplings such that in terms of these couplings the
$\beta$-function is finite (i.e. $N$-independent).  We will denote the
vector of such rescaled couplings of the theory by ${\bf g}_t$. The
statement of finite $\beta$ function is that the RG equation is
\eqn\rgflow{ \dot{ {\bf g}_t } = F_0({\bf g}_t)  } 
without any explicit dependence on $N$. 
Taking into account ${1\over N}$ correction we can expand the $\beta$
function by
\eqn\rgexpnd{ \dot{{\bf g}_t} = F_0({\bf g}_t)+{1\over N} F_1(
{\bf g}_t)+ {1\over N^2} F_2({\bf g}_t)+ \cdots .}  A familiar example
is that of the `t Hooft effective coupling in the large N limit of
QCD. The relevant term in the $\beta$-function at finite N is
$$\dot{(g_{\rm YM}^2)}= b_0 N g_{\rm YM}^4 + \cdots,$$ 
which implies that the
$\beta$-function for the `t Hooft coupling $g_{\rm eff}^2 = 
g^2 N$ is independent of N.

If the classical supergravity vacuum is such that the dual theory at
large N is conformal, viz. the spacetime is of the type $AdS_p\times
M_{11-p}$, then the functional form of $F_0$ is constrained to be
$$F_0({\bf g}_t)={\bf M} \cdot {\bf g}_t, $$ where $\bf M$ is 
a matrix which encodes the dimensions
of the operators around the fixed point at large $N$. We also relabeled
the couplings such that the large $N$ fixed point is at ${\bf g}_t = 0$.

Concentrating on the invariant operators, if there are no marginal
operators, the effect of the ${1\over N}$ corrections to these
operators will be small and controllable. The
reason is that even when we take the ${1\over N}$ corrections into
account there still is a fixed point close by. Expanding $F_1({\bf
g}_t)$ around the large $N$ fixed point, the renormalization group
equation, to leading order in $1
\over N$-correction and ${\bf g}_t$, is
\eqn\newrg{{\dot{{\bf g}}_t}={\bf M} \cdot {\bf g}_t +{1\over N} F_1(0)
+ \cdots}
and is solved by
\eqn\newfp{{\bf g}_t^*=-{1\over N}\,\,{\bf M}^{-1} \cdot F_1(0) + \cdots 
.}  This is possible because under the assumption that there are no
marginal operators the matrix $\bf M$ is invertible. Shifting the
value of the fixed point in terms of the coupling is exactly analogous
to giving small expectation value to non-zero mass fields. 

Under more limited circumstances, the same line of argument can be
applied to the case where the conformal field theory at the large N
limit has marginal operators but not truly marginal ones\foot{We are
indebted to N. Seiberg for discussion of this point.}. For example, if
the large N limit renormalization group flow near ${g}_t=0$ is of
the form
$${\dot{g}_t}= a \, {g}_t^2 + {1\over N} \, b + \cdots, $$
 where $a$ and $b$ are numbers, then there exists a fixed point at 
${g}_t^{*2}=-{1\over N}{b\over a}$ provided ${b\over a}<0$.

We see that this kind of instability in non-supersymmetric theories
occurs only when there are marginal operators and, in some cases, when
there are marginal but no exactly marginal operators. In the case
where in large N there are truely marginal operators then the effect
of $F_1$ on the submanifold\foot{in the space of coupling constants}
of truely marginal deformations is pronounced. Since there is no flow
on this submanifold at leading order in $1/N$, the sub-leading terms
control the flow and there may simply not be a fixed point or it may
be far away from our starting point. This will be the case with any
typical vacua that is derived in string perturbation theory, where the
dilaton will be associated with an invariant marginal operator. We
will therefore discuss M-theory and specific vacua thereof and show
that they do not have any marginal perturbations.

One might similarly obtain four-dimensional conformal field theory
from Type IIB on $AdS_5 \times M^5$, but not in the perturbative
regime. The generic problem, as we have explained above, is the
tadpole for the dilaton field. However, there cannot be such a tadpole
for the dilaton if it is fixed at the self-dual point under
$SL(2,Z)$. In this case, the discrete gauge symmetry prevents
generation of such a tadpole.

\newsec{$d=3$ Non-Supersymmetric Conformal Field Theory and 
$AdS_4 \times {X}_7$}

The most extensive and varied list of M-theory comactification of the
form $AdS_p\times X^{11-p}$ is available for the case $p=4$. A large
class of vacua is known and the stability (in the sense that all
fields satisfy the Breitenlohner-Freedman unitarity bound
\ubound) of many of them has been analyzed. In
particular, it has been shown that there are many non-supersymmetric
yet stable vacua.  To achieve the goal of this paper, we are
interested in examining the question whether these vacua have massless
fields on $AdS_4$ space that are invariant under all the gauge
symmetries of the supergravity theory.  If a specific vacuum does not
contain any massless, gauge-singlet scalar field, one would then
conclude that the vacuum passes the tadpole-hurdle alluded in the
previous section even after the $1 \over N$-corrections are taken into
account. Such a vacuum may be dual to a non-supersymmetric conformal
field theory.

A subset of the non-supersymmetric vacua of the form $AdS_4\times X^7$
is obtained by ``skew-whiffing'' supersymmetric compactification vacua
\duffreview\squash. The advantage of this method is that it guarantees
the stability of the resulting non-supersymmetric vacua \dnp, at
leading N. This is evidently so as the mass spectrum of the bosonic
fields that might potentially cause an instability remains unchanged
from that of the initial supersymmetric vacua and hence is stable. The
disadvantage from our perspectives is that, in the vacua with more
supersymmetries, one often encounters exactly marginal
operators. According to our previous argument such operators, if exist
and are invariant, would cause an instability to the
non-supersymmetric vacua (those obtained by "skew whiffing") once
$1/N$-corrections are made. On the other hand, exactly marginal
operators become scarce for compactifications with less
supersymmetries. We will henceforth focus on non-supersymmetric vacua
obtained from ``skew-whiffing'' of compactifications with {\sl
smaller} supersymmetries.

\subsec{Basics of ``skew-whiffed'' non-supersymmetric 
compactification vacua}

Let us briefly review known facts about the mass spectrum of
``skew-whiffed'' compactification vacua. The ``skew-whiffing''
procedure is to reverse the orientation of the compactification
manifold $X^7$.  This is done by replacing $e^\mu_a$ by $-e^\mu_a$. The
resulting background generated in this way is clearly a solution of the 
supergravity equations of motion.
However, the amount of the residual supersymmetry
preserved by the "skew-whiffed" vacuum changes in general.  Indeed,
apart from the most symmetric choice such as $X^7=S^7$, no
supersymmetry is left preserved after the orientation reversal of $X^7$ 
(in so far as the manifold $X^7$ is smooth).

Starting from the original supersymmetric vacuum, the mass spectrum of
the ``skew-whiffed'' vacuum is obtained by interchanging the negative
and the positive parts of the spectrum of the appropriate Laplacian
operators on the compact manifold $X^7$ \dnp\duffreview . Since the
spectrum of these operators determines the masses of supergravity
fields on $AdS_4$, the masses of some of the fields may change. For
example, the spectrum of the spinors changes, corresponding to the
fact that the new vacuum typically leaves out no residual
supersymmetry at all.

Being non-supersymmetric, one might suspect that the ``skew-whiffed''
vacua generically contain some fields violating the unitarity bound
and develop tachyonic instability. The supergravity fields that can
potentially violate the unitarity bound are actually those with
quantum numbers $J^P = 0^+$ on $AdS_4$. They arise from (traceless)
deformations of the metric field on the compact manifold
$X$. Fortunately, the mass spectrum of these fields does not change
from that of the starting supersymmetric vacuum and hence the
non-supersymmetric vacuum will remain stable as well
\dnp,  at least for large $N$.

\subsec{Generic problems}

The argument given above holds only in the strict $N\rightarrow\infty$
limit.  Clearly, the $1 \over N$-corrections will be different for the
supersymmetric vacua and for their ``skew-whiffed'' ones. In the above
discussion, we have utilized the underlying supersymmetry to argue
that both the starting vacuum and the ``skew-whiffed'' one are stable
at large N. However, the very same supersymmetry may also be a source
of potentially dangerous operators.  One needs to be aware of that the
following problems might arise:

\medskip

\noindent 1. {\tt fields saturating unitarity bound may be present}: 
Since the $J^P = 0^+$ part of mass spectrum is the same as in the
starting supersymmetric compactification, there might be fields that
saturate the unitarity bound (and corresponding operator with
dimensions-$3/2$ in the dual conformal field theory). The reason for
this is that, in highly supersymmetric field theories, such operators
occur quite commonly (for example, in three-dimensional ${\cal N}=8$
gauge theory, the operator $tr X^{\{i}X^jX^{k\}}$ has the scaling
dimension $3/2$). To avoid this problem, we will consider
compactifications with less supersymmetries, for which generically we
do not expect to find such operators. We then ``skew-whiff'' these
compactifications and begin with the resulting non-supersymmetric
vacua.

\medskip

\noindent 2. {\tt (exactly) marginal operators may be present}: 
Since many of the dimensions in the ``skew-whiffed'' compactifications
are inherited from the supersymmetric ones, there might be massless
particles in the spectrum, inherited from exactly marginal operators
in the original supersymmetric ones. As before, in supersymmetric
gauge theories, it is quite common that (exactly) marginal scalar
operators are present.  Our main concern is whether a given
compactification vacuum has any massless fields invariant under all
supergravity gauge symmetries. While certainly a much more restricted
class, it is again advantageous to focus on ``skew-whiffing'' of
compactifications with less supersymmetries as they would not have
exactly marginal operators in general.

\medskip

Before we proceed to specific examples, let us identify which modes of
the scalar field might be potentially dangerous in the sense that they
will correspond to massless gauge singlet fields on the $AdS_4$.
Since our foregoing discussion will be based on results from
\duffreview, we first note that the definition of mass in
\duffreview\ is shifted from that we will be using momentarily. In
\duffreview , the mass spectrum is defined according to $ (\triangle -
8m^2 + M^2 ) S=0$ (Eq.(3.2.22) in \duffreview ), where $\triangle$ is
the scalar Laplacian on the $AdS_4$ space, $m^2$ is a parameter
associated with the compactification, and $M^2$ is the mass.  In what
follows, we will define the mass $\tilde M$ via $( \triangle + {\tilde
M}^2) S=0$, viz., $${\tilde M}^2= M^2-8 m^2. $$

We are interested in identifying possible massless modes. The modes
that are potentially dangerous to develop tadpoles are those of $J^P =
0^+$.  Those of $J^P = 0^-$ cannot develop a tadpole as such a tadpole
would break parity, viz., the symmetry $x^1\rightarrow - x^1,\
C\rightarrow -C$, where $x^1$ is one of the spatial coordinates in
$AdS_4$ and $C$ is the three-form potential of M-theory. Under our
assumption of {\sl calculability}, we expect this to be a symmetry of
the dual conformal field theory.  This symmetry then forbids
generation of the tadpoles for the parity-odd fields.

Hence, potentially dangerous modes that might lead to massless states are
$$ 0^{+(1)} : \hskip0.3cm 
\triangle_0+36m^2-12m^2 (\triangle_0+9m^2)^{1\over2} $$
$$ 0^{+(2)} : \hskip0.3cm \triangle_L-12m^2. \hskip3.15cm $$ 
The first will give rise to a dangerous mode
for $\triangle_0=72m^2$ ($\triangle_0=0$ is omitted from the spectrum)
and the second will give a dangerous mode for $\triangle_L=12m^2$.

\subsec{Examples}

We will now present three examples of compactification of the form
$AdS_4\times X^7$, which are free from massless, gauge-singlet scalar
fields.  The full spectrum is known in the literature only for the
first case.  For the other two, the full spectrum is not known. As
such, it is not clear whether there will be a field that saturates the
unitarity bound.  However since these will the ``skew-shiffed''
vacuum of a lower supersymmetry compactification, we do not expect
such a scalar field in such compactifications to be present.

\vskip0.5cm
{\it Example 1: $AdS_4\times S^7/Z_2$}

There are two types of $AdS_4 \times S^7/Z_2$ compactifications (even
before taking into account different torsion classes \wtntor\savetc)
related by ``skew-whiffing''. M-theory on $S^7$ is dual to
three-dimensional ${\cal N}=8$ $SU(N)$ super-Yang Mills theory
\adsfour.  The supersymmetry charges would transform in one of the two
spinor representations of $SO(8)$, say, $8_s$. Since we take a quotient
by $Z_2$ which acts as $-1$ on the $8_v$ spinor, we can lift it to act
as $+1$ on the $8_s$ spinor, in which case we still have ${\cal N}=8$
supersymmetry. Alternatively, if it is lifted to act as $-1$ on the
$8_s$ spinor, then supersymmetry is broken completely.  Even though we
have not casted it this way, this is basically equivalent to the
``skew-whiffing'' process.

It is easy to see that this compactification has no scalar field that
can violate the unitarity bound. The only scalar field that lies at
the unitarity bound \adsfour\ for $S^7$ is $tr X^{\{i}X^jX^{k\}}$, but
it is projected out by $Z_2$ quotient.  One can also verify that there
is no scalar field that can generate dangerous tadpoles: the only
marginal operator with the $J^P = 0^+$ quantum number is in the symmetric 
traceless 6-tensor representation of $SO(8)$.

\vskip0.5cm
{\it Example 2: The Squashed 7-Sphere}

It is known that there is one supersymmetric and one
non-supersymmetric squashed 7-spheres \squash. The supersymmetric one
has four-dimensional ${\cal N}=1$ supersymmetry. The spectrum of
Laplacians relevant for the $J^P = 0^+$ states is the same in both
cases and is given by Eqs.  (8.4.2), (8.4.9) and (8.4.10) in
\duffreview:
$$ \triangle_0 = {20\over 9} m^2 C_G \hskip8cm $$
$$ \triangle _L = {20\over 9} m^2 \left(C_G+{9\over5} \right) \ \ 
\ \ {\rm or}
\ \ \ \ {20\over 9} m^2 \left(C_G+{8 \over 5} \pm {2 \over \sqrt{5}}
\left(C_G+{1 \over 20} \right)^{1\over2} \right), $$
where $C_G=C_{SO(5)}+3C_{SU(2)}$ .
Since we interested in gauge-singlet fields, $C_G=0$ and the masses
of the relevant $J^P = 0^+$ obtained in this way turn out always non-zero.

\vskip0.5cm
{\it Example 3: $N(k,l)$}


The manifold $N(k,l)$ is defined \casta\ as a coset $[SU(3)\times
U(1)]/ [U (1)\times U (1)] $. The integers $k$ and $l$ parameterize
different embeddings of the two $U(1)$'s into the maximal torus of
$[SU(3)\times U(1)]$.  For more details, the reader is referred to
\casta. The supersymmetric compactifications on $N(k,l)$ have 
${\cal N}=1$ 
on $AdS_4$, and we will be interested in the "skew-whiffed"
counterpart of them.

To examine whether there can be $SU(3)\times U(1)$ invariant massless
field that might destabilize the vacuum, we again check possible 
massless fields in the spectrum of $ J^P = 0^{+(1)}$ and $J^P = 0^{+(2)}$.
For a symmetric space, however, $J^P = 0^{+(1)}$ does not lead to any
dangerous modes. The reason is that these modes originate from the
supergravity fields that are scalars on $X^7$ (hence eigenstates of the 
scalar Laplacian), and the only invariant mode is the constant mode for 
which $\Delta_0=0$.  Such a constant mode, however, is not physical.

The analysis of the $0^{+(2)}$ mode is also straightforward. We are
again interested only in modes which are invariant under $SU(3)\times
U(1)$. The analysis of these modes is discussed in full detail in
\casta. The fact that there is no invariant massless scalar field is 
essentially encoded in the computations of \casta, where the authors find
only a unique solution for every value of $k$ and $l$. Nevertheless,
for completeness we briefly outline here the computation of the
determinant of the mass matrix for these fluctuations. The result
would be non-zero, indicating that there are no massless invariant
fields.

To analyze the allowed metrics that preserve the $G=SU(3)\times U(1)$
symmetry, we first decompose the adjoint representation of this group 
into irreducible representations under the $ad(U(1)\times U(1))$. 
After we discard the
subspace of $G$ which generates H, the remaining linear space is
tangent to $X^7$ at the H-orbit that passes through the identity
of $G$. The metric of this space splits into a sum of metrics on each of 
the $U(1)\times U(1)$ invariant subspaces. The freedom that we now have is
to multiply each such component by an arbitrary number.

For $N(k, l) = [SU(3)\times U(1)]/ [U(1)\times U(1)]$, this tangent
 space splits into four irreducible representations, three of them are
 of real-dimension 2 and one which is of real-dimension 1.  These will
 be parameterized by indices $(a,b,..),(A,B,..), ({\dot A},{\dot
 B},..)$ and $Z$, respectively. Denoting by $g^0$ the G-invariant
 metric (on $ad(g)$) the invariant metric on $M_7$ is of the form:
\eqn\mtrc{g_{ab}=\alpha g^0_{ab},\ \ \ 
  g_{ZZ}=\beta g^0_{ZZ},\ \ \ g_{AB}=\gamma g^0_{AB},\ \ \
 g_{{\dot A}{\dot B}}=\delta g^0_{{\dot A}{\dot B}}.}  
The solution to Einstein's
  equations is given in \casta, where it is written in terms of the
  variables
$$a={\alpha^2\over \delta^2},\ \ \ \ b={\alpha^2\over\gamma^2},\ \ \ \ u=
+ {\alpha\gamma\over\beta\delta}{3p+q\over \sqrt{2}},\ \ \ \ v=
- {\alpha\delta\over\beta\gamma}{3p+q\over \sqrt{2}}.$$ The solution is
unique.  

Once we have found the solution
  of Einstein equations, the allowed G-invariant fluctuations are
  fluctuations of $\alpha$, $\beta$, $\gamma$ and $\delta$ subject to
  the constraint that the total volume is invariant
$$2\va+\vb+2\vc+2\vd=0.$$ Under small fluctuations of the three 
independent 
scale parameters $\alpha$, $\gamma$ and $\delta$, we find that 
\eqn\vrab{ 
\delta R^a_b = {\cc^2\dd^2\over \aa^2}
\left[ {3\over 2}ab+{1\over 4}(-1-a^2-b^2)-(av+bu)^2 \right]\va
\hskip4.5cm }
$$+ \left[{1\over4}(1+b^2-a^2)-{1\over2}(av+bu)^2 \right] \vc 
+ \left[ {1\over4} (1-b^2+a^2)-{1\over2}(av+bu)^2 \right] \vd$$
\eqn\vrzz{
\delta R^Z_Z= {\cc^2\dd^2\over\aa^2} 
 \left[(av+bu)^2+{1\over 2}u^2+{1\over 2}v^2 \right]\va \hskip5.6cm}
$$ + {\cc^2\dd^2\over\aa^2}
 \left[ {1\over2}(av+bu)^2 +{1\over 2}v^2+u^2 \right] \vc
+ {\cc^2\dd^2\over\aa^2}
 \left[ {1\over2}(av+bu)^2+v^2+{1\over 2}u^2 \right] \vd$$
\eqn\vrbabb{
\delta R^A_B={\cc^2\dd^2\over\aa^2} \left[
{1\over4}(b^2+1-a^2)-{1\over2}u^2  \right] \va \hskip5.9cm}
$$+ \left[{3\over2}a+{1\over4}(-b^2-1-a^2)-u^2 \right] \vc+
\left[ {1\over4}(b^2-1+a^2)-{1\over2}u^2 \right] \vd$$
\eqn\vrdadb{ \delta R^{\dot A}_{\dot B} = {\cc^2\dd^2\over\aa^2} 
\left[ {1\over4}(a^2+1-b^2)-{1\over2}v^2 \right] \va \hskip5.5cm}
$$ + \left[ {1\over4}(a^2-1+b^2)-{1\over2}v^2 \right] \vc
+  \left[{3\over2}b+{1\over4}(-a^2-1-b^2)-v^2 \right] \vd, $$
where only three out of the four variational equations are 
actually independent.

Choosing the three independent equations appropriately, we can cast 
these equations into a form 
$$\delta R_\mu^\mu ={M^\mu}_i {\Delta V_i\over V_i}$$ where
$\mu=a,A,{\dot A}$ and $V_i=\alpha,\gamma,\delta$. From these
expressions, we can deduce the equations of motion for the fluctuations
$\Delta\alpha,\Delta\gamma$ and $\Delta\delta$. The result is that the
mass matrix is the matrix $M^\mu_i$, up to multiplications by a
non-degenerate matrix. Since we are interested in zero modes, this will
not matter since we can evaluate the determinant of $M$, using the
values of $\alpha,\gamma,\delta$ of the solution in \casta. The result
is that there are no zero mass fields, i.e., there are no massless
invariant fields that can acquire dangerous tadpoles.

Finally, a remark is in order. One might also prompt to explore stable
higher-dimensional non-supersymmetric compactifications, especially,
of the form $AdS_7 \times X^4$ and the corresponding dual conformal
field theories. In the previous version of this letter, we have
indicated $AdS_7 \times S^2 \times S^2$ might be a possible
non-supersymmetric compactification that enables us to address the
stability along the lines of section 2. It turned out that this vacuum
is actually not a good starting point for $1 \over N$-expansion: the
compactification is unstable under fluctuations for which the volume
of the first $S^2$ shrinks and that of the second $S^2$ expands while
keeping the total volume of $S^2 \times S^2$ fixed. Hence, barring the
possibility of orbifold construction, it then appears that there is no
stable non-supersymmetric compactification to $AdS_7$.

\vskip 1cm

\centerline{\bf Acknowledgments}\nobreak

We would like to thank O. Aharony, J. Distler, S. de Alwis, M.J. Duff,
O. Ganor, S. Gubser, A. Kapustin, I. Klebanov, N. Seiberg and
F. Zamora for illuminating discussions. The work of MB is supported by
NSF grant NSF PHY-9513835. The work of SJR is supported by KOSEF
Interdisciplinary Research Grant and SRC-Program, KRF International
Collaboration Grant, Ministry of Education BSRI 98-2418 Grant, SNU
Research Fund, and The Korea Foundation for Advanced Studies Faculty
Fellowship.

\listrefs

\end